\documentclass[a4paper,10pt,twoside]{cpc-hepnp}

\usepackage{fancyhdr}            
\usepackage{multicol}            
\usepackage{upgreek}             
\usepackage{indentfirst}         
\usepackage{booktabs}            
\usepackage{array,tabularx}      
\usepackage{keyval,graphicx}     
\usepackage{textcomp}            
\usepackage{amssymb,bm,mathrsfs,bbm,amscd} 
\usepackage[tbtags]{amsmath}     
\usepackage{lastpage}            
\usepackage[numbers,sort&compress]{natbib}  
\usepackage{color}

\begin{document}

\setlength{\abovecaptionskip}{4pt plus1pt minus1pt}   
\setlength{\belowcaptionskip}{4pt plus1pt minus1pt}   
\setlength{\abovedisplayskip}{6pt plus1pt minus1pt}   
\setlength{\belowdisplayskip}{6pt plus1pt minus1pt}   
\addtolength{\thinmuskip}{-1mu}            
\addtolength{\medmuskip}{-2mu}             
\addtolength{\thickmuskip}{-2mu}           
\setlength{\belowrulesep}{0pt}          
\setlength{\aboverulesep}{0pt}          
\setlength{\arraycolsep}{2pt}           

\providecommand{\e}[1]{\ensuremath{\times 10^{#1}}}


\fancyhead[c]{\small Chinese Physics C~~~Vol. 37, No. 5 (2013) 056002}
\fancyfoot[C]{\small 056002-\thepage}

\footnotetext[0]{Received 26 June 2012}

\title{\boldmath A time-of-flight system for {the} external target facility\thanks{Supported by National
Natural Science Foundation of China (11205210)}}

\author{%
ZHANG Xue-Heng$^{1;1)}$\email{zhxh\oa impcas.ac.cn}
\quad YU Yu-Hong$^{1}$
\quad SUN Zhi-Yu$^{1}$
\quad MAO Rui-Shi$^{1}$\\
WANG Shi-Tao$^{1}$
\quad ZHOU Yong$^{1, 2}$
\quad YAN Duo$^{1,2}$
\quad LIU Long-Xiang$^{1,2}$%
}

\maketitle

\address{%
$^1$ Institute of Modern Physics, Chinese Academy of Sciences, Lanzhou 730000, China\\
$^2$ Graduate University of Chinese Academy of Sciences, Beijing 100049, China\\
}

\begin{abstract}
A time-of-flight system with {a} plastic scintillator coupled
to photomultipliers {is} developed for {the} external target facility (ETF).
This system can satisfy the requirement of {an} ultrahigh vacuum ($\sim
$10$^{-9}$ mbar), {a} high counting rate ($\sim $10$^{6}$ particles per second)
and {a} magnetic field environment. In the beam test experiment, {a} total time
resolution of 580 ps FWHM was obtained for the whole system, and {nuclei}
with {a} mass {of} up to 80 could be identified {using} this system.
\end{abstract}


\begin{pacs}
29.40.Mc \qquad {\bf DOI:} 10.1088/1674-1137/37/5/056002
\end{pacs}

\footnotetext[0]{\hspace*{-3mm}\raisebox{0.3ex}{$\scriptstyle\copyright$}2013
Chinese Physical Society and the Institute of High Energy Physics
of the Chinese Academy of Sciences and the Institute
of Modern Physics of the Chinese Academy of Sciences and IOP Publishing Ltd}%

\begin{multicols}{2}

\section{Introduction}

An external target facility (ETF) is constructed downstream {of} the second
radioactive ion beam line {(RIBLL2)} in the Heavy Ion Research
Facility in Lanzhou (HIRFL) [1, 2]. It consists of a series of sub-detector
systems, such as {the} $\gamma $ ball [3], TOF wall [4], neutron wall [5], and
MWPCs [6]. Using {beams} that can {accelerate} nuclei up to $^{238}$U
with a kinetic energy of several {hundreds of} MeV to GeV {using} the main Cooling
Storage Ring (CSRm) [2], {research} on the structure of exotic nuclei
and the equation of state of dense nuclear matter can be carried out at {the} ETF.

Like the FRS (GSI FRagment Separator), which delivers radioactive ion beams
(RIBs) to the ALADIN-LAND setup for decay and reaction studies at GSI [7,
8], the first half of RIBLL2 can be used to produce and separate the
interested RIBs for ETF. The RIBs can be identified by combining the
time-of-flight (TOF), the energy deposit $\Delta E$ and the magnetic
rigidity $B\rho $, which is widely used by all projectile fragmentation
type separators. Therefore, a TOF system, which provides experimental
trigger and particle identification, should be developed for {the} ETF. This
system should satisfy the following stringent {requirements}: (\ref{eq1}) the start
detector should have a large area ($\sim $100~mm$\times $100~mm),
{be} used in the ultrahigh vacuum ($\sim $10$^{-9 }$~mb), and should
withstand a high counting rate ($\sim $10$^{6}$ particles per second); (2)
{the} stop detector should be operated in the magnetic field environment
($<$0.01 T); and (3) the TOF system should have a good mass resolution power
(up\linebreak to 60).

In this paper, we would like to report the design of this TOF system. The
{measurements results} will also be presented.

\section{The design of the TOF system}

The layout of the first half of {the} RIBLL2 and ETF is shown in Fig.~1. The
primary beam from CSRm hits a target at {the} F0 cave. The RIBs produced by
projectile fragmentation are separated and purified by the combined
$B\rho$-$\Delta E $-$B\rho$ method [9], and then delivered to {the} ETF for experimental
studies. Particle identification is implemented before the second target. To
obtain a good mass resolution, the flight path should be as long as
possible, {so} we place the start and stop detectors at the F1 cave and
upstream of the second target, respectively. The total length of the flight
path is about 26~m.

\subsection{The start detector}

The start detector is designed based on {a} plastic scintillator from the
manufacturer Eljen (EJ200) [10]. Hamamatsu R7111 photomultipliers [11] are
used to read out the signals from both ends of the scintillator.

According to the beam {optics} calculations [12], the beam spots have maximum
position dispersion at {the} F1 cave. The beam size is about 100~mm$\times $100~mm
when the slits are fully opened. {The} plastic scintillator, which has
the feature of {no} size {restrictions}, can be easily made into {a} large area detector
{to} meet the size requirement.

\end{multicols}
\begin{center}
\includegraphics{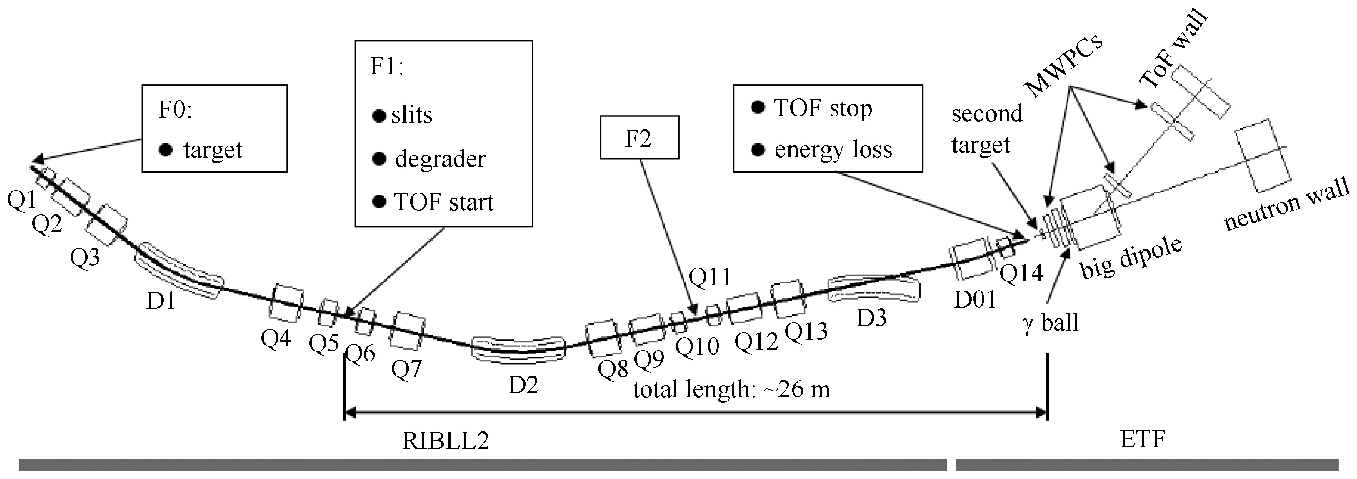}
\figcaption{Top view of the first half of {the} RIBLL2 and ETF.}
\end{center}
\begin{multicols}{2}

\end{multicols}
\begin{center}
\includegraphics{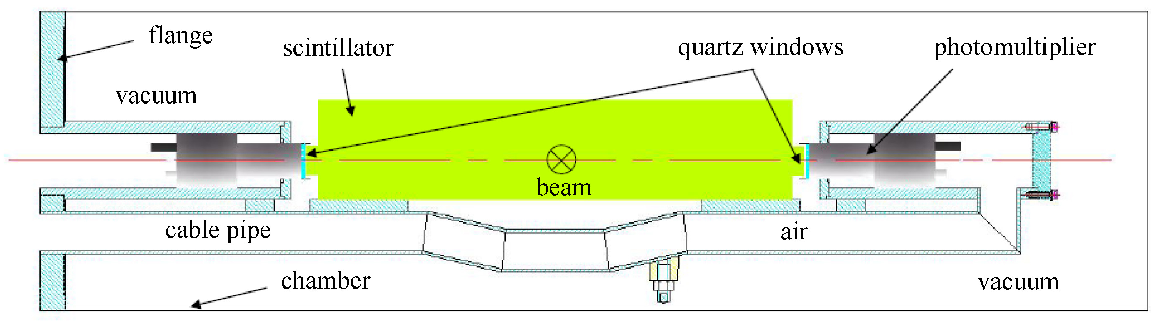}
\figcaption{(color online) Schematic layout of the start detector.}
\end{center}
\begin{multicols}{2}

The start detector will be used in ultrahigh vacuum (at {a} pressure of
$\sim$10$^{-9 }$~mb). This calls for very low outgassing materials in {the} mechanical
support, optical coupling, {a} voltage divider and cables. A specific stainless
steel frame has been designed and its schematic layout is shown in Fig.~2.
The whole frame is fixed on a flange connected with the F1 chamber. The
scintillator is placed in the center of the beam line, and photomultipliers
are fixed in the cylinders at both ends of the scintillator. Two 3~mm thick
circular quartz windows are welded at one end of two cylinders to isolate
the air and to transmit photons from the scintillator to {the} photomultipliers.
The {couplings} via optical silicon pads and {a} vacuum are chosen between
{the} photomultipliers and quartz windows, and between {the} quartz windows and
scintillator, respectively. With this frame all the materials except the
scintillator are placed in the atmosphere to ensure that the start detector
is less affected by {the} outgassing rate.

To ensure the air tightness of {the} welded interfaces, the quartz windows can
only be processed into a circle with a maximum diameter of 35~mm due to the
{constraints} of {the} machining technology. The light collection efficiency will be
influenced by the difference {in} sizes between the scintillator and the
quartz windows. The light collection efficiency of two prototypes has been
simulated with GEANT4 [13]. Prototype I, shown in Fig.~3(a), consists of a
scintillator {of} dimension 100~mm$\times $100~mm$\times $3~mm and two {vacuum-coupled}
light guides. Prototype Ⅱ\/, shown in Fig.~3(b), is a
monoblock scintillator without four corners. The thickness of Propotype Ⅱ\/
is 3~mm, and the widths of the cut and uncut parts are 30~mm and 100~mm,
respectively. The total length of {the} two prototypes, which is equal to the
distance of two quartz windows, is 500~mm. In the simulation both
{the} scintillator and light guides are covered by thin aluminum foil, {the}
refraction of {the} scintillator and light guide is 1.58, and the vacuum
refraction is 1.0. A primary beam of 500~MeV/u $^{ 12}$C with a spot size of
80~mm in diameter hits the center of the scintillator. The photons will be
produced by the scintillation process due to the electromagnetic energy loss
of $^{12}$C. The light collection efficiency is calculated by the ratio
between the photons transmitted to the end ($N_1$) and the {total} photons
($N_0$). Fig.~3(c) shows the light collection efficiency as a function of the
length of {the} vacuum layers between the scintillator and light guides for
Prototype Ⅰ, and Fig.~3(d) shows the light collection efficiency change with
the length of the cut corners for Prototype Ⅱ. It can be seen that the
light collection efficiency slightly increases with {a} decrease {in} the length
of {the} cut corners, and the value is similar to the efficiency at the vacuum
length of 1~mm. Considering the accuracy of machining, Prototype Ⅱ\/ is
{finally} adopted with a cut corner length of 6~mm. The scintillator is
covered by one layer of 6~$\upmu $m thick aluminum foil, and the length
between the scintillator and the quartz window is about 1~mm.

When the start detector is running at high rate, the

\end{multicols}
\begin{center}
\includegraphics{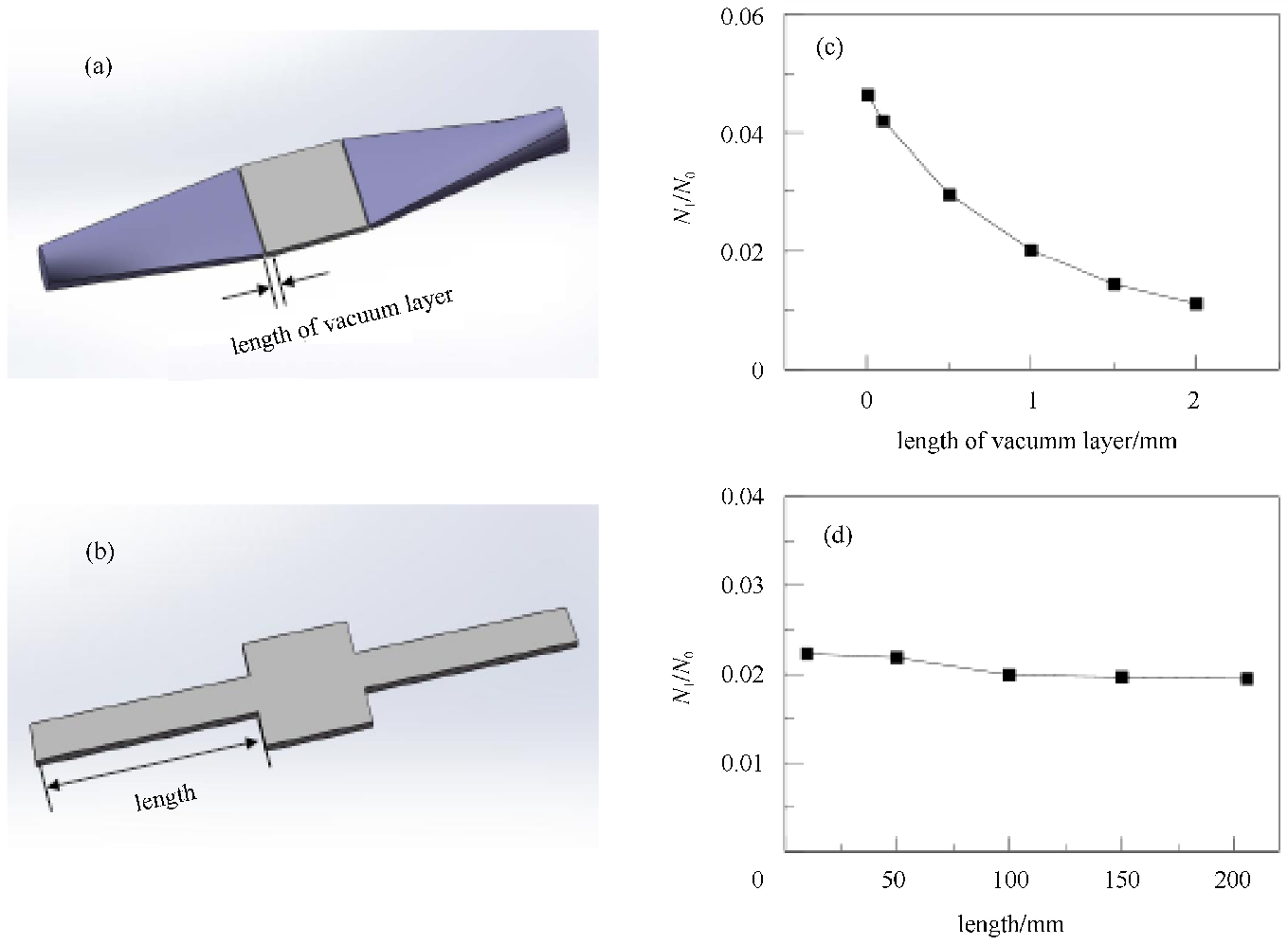}
\figcaption{(color online) Light collection efficiency {simulated} with
GEANT4. (a) and (b) are a schematic diagram of {prototypes} Ⅰ\/ and Ⅱ, (c)
is the efficiency as a function of the length of {the} vacuum layer of Prototype Ⅰ,
and (d) shows the efficiency as a function of the length of cut corners of Prototype Ⅱ.}
\end{center}
\vspace{-5mm}
\begin{multicols}{2}

\noindent anode current of {the}
photomultipliers increases dramatically, and the non-linearity effect
appears in such a way that the voltage drop in the last amplification stages
becomes significant and causes gain instability due to insufficient bleeder
current supply [14].~To keep the photomultiplier gain stable against the
incoming particle flux (up to $\sim $10$^{6}$ particles per second), three
types of voltage dividers are designed, as shown in Fig.~4(a),\linebreak (b) and (c),
respectively. Type Ⅰ\/ is a traditional voltage divider. Type Ⅱ\/ is an
improved voltage divider with a large-scale regulated power supply network,
and Type Ⅲ\/ adopts the additional power supplies at {the} last three dynodes
based on Type Ⅱ. The divider ratios of {the} three types of voltage dividers have
been adjusted to obtain good gain [15]. In order to evaluate the rate
capability of our prototypes, a test was performed. A green LED (light
emitting diode) source [16], which was driven by the square signal with
an amplitude of 5~V, width of 40~ns, leading edge of 13~ns and trailing edge of
12.1~ns, was used to simulate an increasing particle rate. {Using} an optical
fiber, the light output was delivered to the center of the R7111
photomultiplier with a high voltage of $-$900 V. {The} amplitude of the
photomultiplier output signal {directly} measured by {an} oscilloscope was studied as
a function of the LED frequency. The results are shown in Fig.~4(d). When the
frequency is increased, the amplitude remains constant up to a maximum value
$\sim $10$^{4}$ Hz for the three voltage dividers, and then the amplitude
increases first and then reduces. The reason {for} the amplitude increase can
be explained as the overlap of the LED electroluminescence. From Fig.~4(d) we
can see that {the} Type Ⅲ\/ voltage divider can still keep the photomultiplier
gain stable at about 10$^{6}$ Hz. So {the} Type Ⅲ\/ voltage divider is used in our detector.

The start detector has been tested with {a} $^{60}$Co gamma ray source, and the
coincidence signals of both ends of the detector measured with {an} oscilloscope
are shown in Fig.~5(a), (b) and (c) when the source was put at the center,
{left} and {right} positions of the scintillator, respectively. It can
be seen that the time and amplitudes of two coincidence signals are similar
when the source is put at the center. When the source is put at one side,
the signal from this side is earlier and bigger than the one from the other side.

\subsection{The stop detector}

The stop detector will be placed near the second target. {This} is a focus
point, and the beam size is smaller than the one on the start detector. So a
50~mm$\times $ 50~mm$\times $1~mm BC408 scintillator sheet [17] coupled to
two photomultipliers\, at\, both\, ends\, with\, optical glue is

\end{multicols}
\begin{multicols}{2}

\noindent used for the stop
detector. Since the position of the stop detector is near the big dipole of {the}
ETF, the leakage magnetic flux has a serious influence on the signal
amplitude of {the} photomultiplier. Therefore, {the} R7111 photomultipliers have been
replaced by {fine} mesh R7761 photomultipliers [11], which can be operated
even in strong magnetic fields over 1.0 T [18, 19]. {A} large-scale regulated
power supply network and additional power supplies {for the} last three dynodes
{were} also applied for the {R7761} voltage divider.

\begin{center}
\includegraphics[scale=0.85]{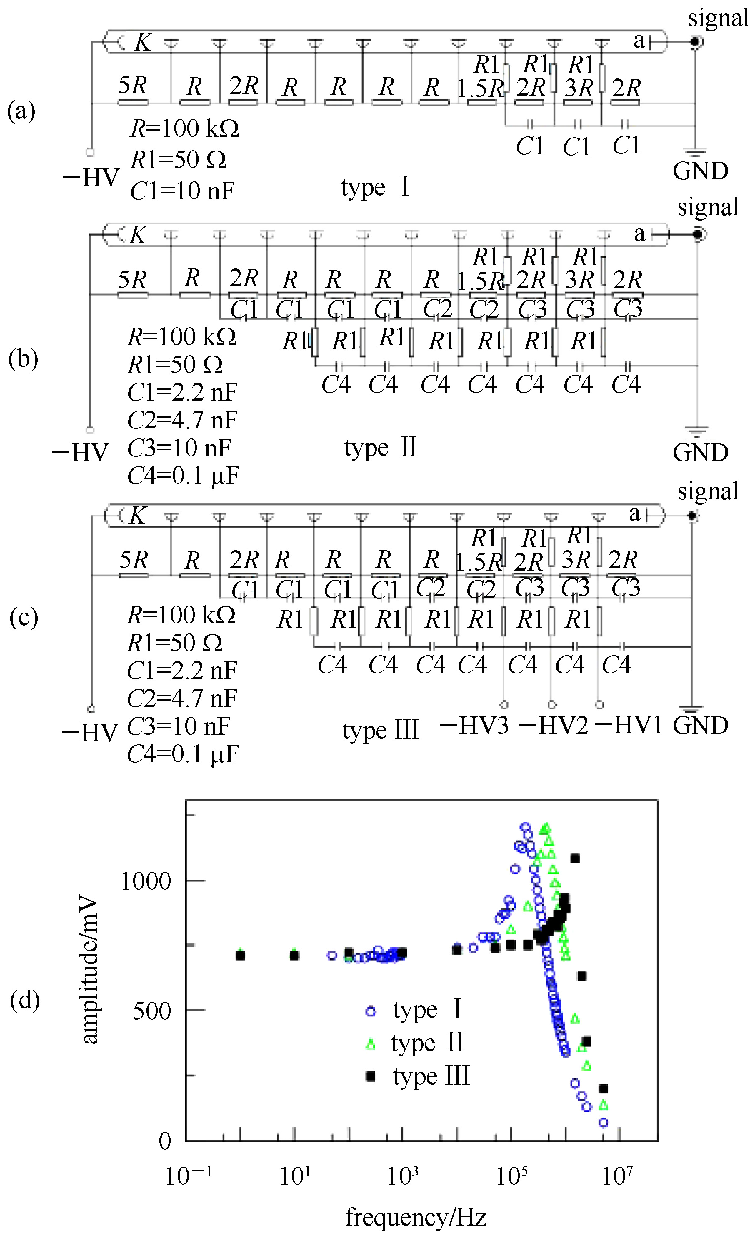}
\figcaption{(color online) Three types of voltage dividers for the R7111
photomultiplier {shown} in (a), (b) and (c), respectively. The signal
amplitudes as a function of LED frequency are shown in (d).}
\end{center}

\section{The {beam test}}

The TOF system {was} installed and used {experimentally}. A primary beam
of 400~MeV/u $^{12}$C was extracted from CSRm and implanted into a 15.1~mm
thick $^{9}$Be target at {the} F0 cave of RIBLL2. The fragmentation products were
separated and identified by the first half of RIBLL2. In the experiment, the
magnetic rigidity was set to 4.57~Tm, which was calculated with $^{9}$C
nuclei. The slits located at {the} F1 cave were opened {at} $\pm $10~mm, and the
degrader was not used.

\begin{center}
\includegraphics[scale=0.8]{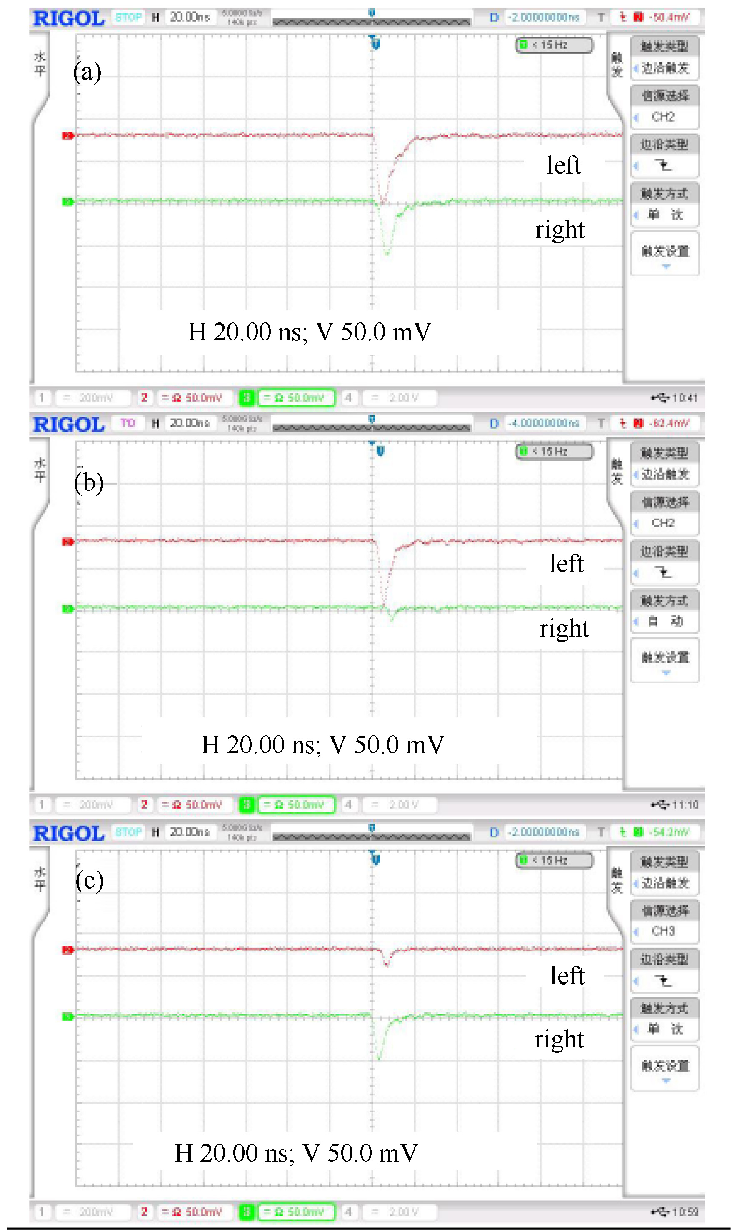}
\figcaption{(color online) Coincidence signals of the start detector measured with
{a} $^{60}$Co gamma ray source. (a), (b) and (c) show the results when the
source was put at the center, {left and right} of the scintillator, {respectively}.}
\end{center}

A typical $\Delta E$-TOF spectrum is shown in Fig.~6(a). The energy loss was
measured by a 140~$\upmu $m thick 45~mm$\times $45~mm silicon detector.
Fig.~6(b) shows a TOF spectrum only for {the} $^{9}$C nuclei chosen in Fig.~6(a).
The root-mean-square width of this distribution is determined to be 2.92.
Considering the TDC channel width of 84.7~ps [20], a time resolution of 580
ps (full width at half maximum, FWHM) was obtained for the whole system.
This value includes all the factors that cause time dispersion, such as
fragment energy dispersion and electronics.

The mass resolution power of {the} TOF system is given by the following
relationships [21],
\begin{equation}
\label{eq1}
R=\frac{A}{\Delta A}=\frac{T}{\Delta T},
\end{equation}
where $A$ and $T$ are the mass and {time-of-flight} of {the} nuclei, and $\Delta A$ and
$\Delta T$ are the widths of {the} mass peak and time-of-flight peak,
respectively. With the RIBLL2 maximum magnetic rigidity of 10.64~Tm [2], the
highest velocity $\beta _{\rm max}$ is about 0.915 for the ion with the mass
to charge ratio of 1.5. From Eq.~(\ref{eq1}) we obtain the mass resolution power
$R\leqslant $160, and {nuclei} with {a} mass {of} up to 80 could be identified using this TOF system.

\end{multicols}
\begin{center}
\includegraphics[scale=0.85]{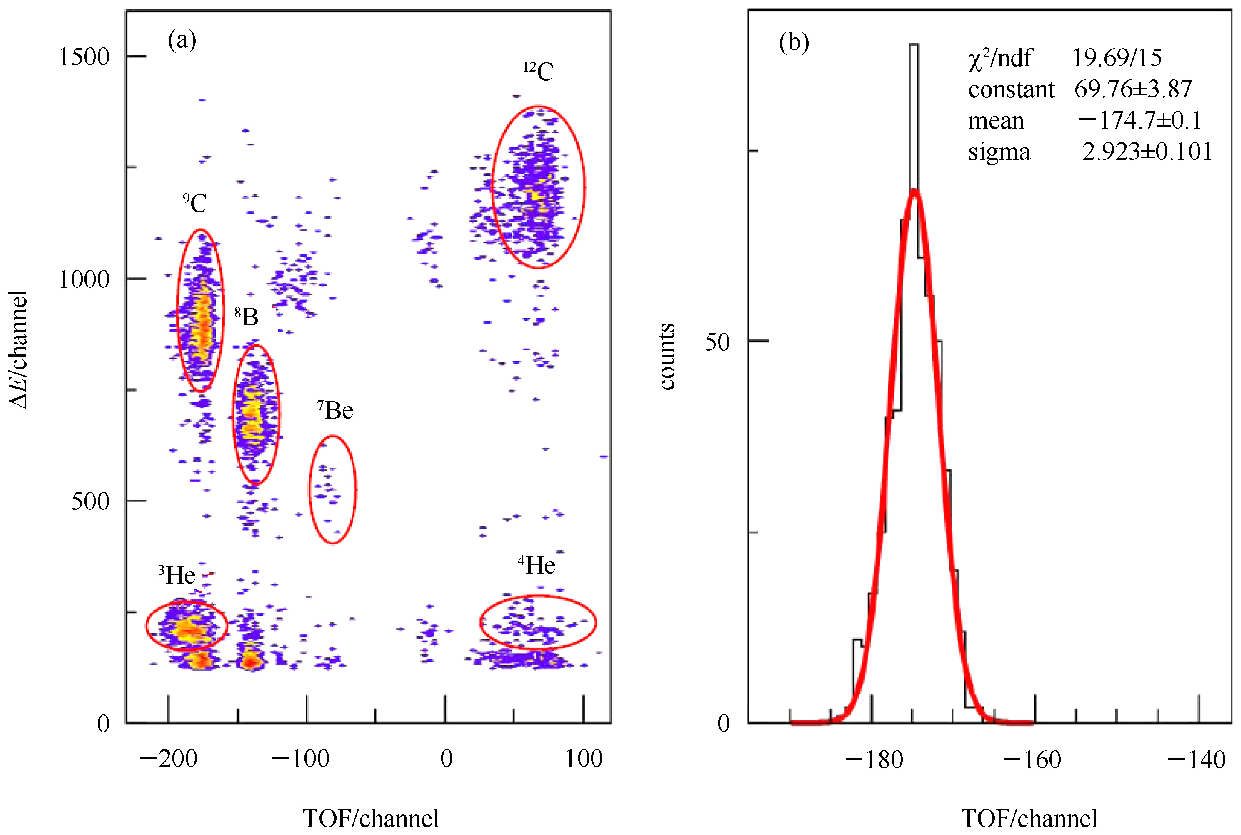}
\figcaption{(color online) {The} particle identification spectrum for $^{12}$C+$^{9}$Be
at 400~MeV/u {(a)}. The TOF spectrum of $^{9}$C is shown in (b),
and the red line denotes the fitting results with Gaussian function.}
\end{center}
\vspace{-4mm}
\begin{multicols}{2}

\section{Summary}

In this paper, a TOF system with {a} plastic scintillator coupled to
photomultipliers {was} developed for {the} ETF. The start and stop detectors
are installed at {the} F1 cave of RIBLL2 and upstream of the second target,
respectively. The distance of this TOF system is about 26~m. The special
design of the frame makes the start detector work in ultrahigh vacuum. To
obtain high light collection efficiency, a monoblock scintillator without
four corners was adopted by the simulations. Photomultipliers for {a} high
magnetic environment were used in the stop detector to resist the leakage
magnetic flux of the big dipole. The voltage dividers for both detectors
were designed as {a} large-scale regulated power supply network {with} additional
power supplies at {the} last three dynodes, which makes the photomultiplier gain
stable at incoming particle flux up to $\sim 10^{6}$ particles per second.
This TOF system was used {experimentally} and a total time resolution of
580~ps FWHM was obtained. {Nuclei} with {a} mass {of} up to 80 could
be identified {using} this system.\\[-2mm]

\acknowledgments{The authors gratefully acknowledge Meng Jun-Hou Shen-Jun and Zhao Yu-Gang
for their support on {the} vacuum technique.}

\end{multicols}

\vspace{-4mm}
\centerline{\rule{80mm}{0.1pt}}
\vspace{-1mm}

\begin{multicols}{2}


\end{multicols}

\clearpage

\end{document}